\documentclass[[aps,prb,reprint,superscriptaddress]{revtex4-1}
\usepackage{amsmath}
\usepackage{bm}
\usepackage{amssymb}
\usepackage{color}
\usepackage{graphicx}
\graphicspath{{pic/}{./}}

\begin{document}

\title{Broadband highly-efficient dielectric metadevices for polarization control}

\author{Sergey Kruk}
\affiliation{Nonlinear Physics Centre, Australian National University, Canberra ACT 2601, Australia}

\author{Ben Hopkins}
\affiliation{Nonlinear Physics Centre, Australian National University, Canberra ACT 2601, Australia}

\author{Ivan Kravchenko}
\affiliation{Center for Nanophase Materials Sciences, Oak Ridge National Laboratory, Oak Ridge, TN 37831, USA}

\author{Andrey Miroshnichenko}
\affiliation{Nonlinear Physics Centre, Australian National University, Canberra ACT 2601, Australia}

\author{Dragomir N. Neshev}
\affiliation{Nonlinear Physics Centre, Australian National University, Canberra ACT 2601, Australia}

\author{Yuri S. Kivshar}
\affiliation{Nonlinear Physics Centre, Australian National University, Canberra ACT 2601, Australia}

\email{sergey.kruk@anu.edu.au}

\begin{abstract}
Metadevices based on dielectric nanostructured surfaces with both electric and magnetic Mie-type resonances have resulted in the best efficiency to date for functional flat optics with only one disadvantage: a narrow operational bandwidth. Here we experimentally demonstrate {\em broadband} transparent all-dielectric metasurfaces for highly efficient polarization manipulation. We utilize the {\em generalized Huygens principle}, with a superposition of the scattering contributions from several electric and magnetic multipolar modes of the constituent meta-atoms, to achieve destructive interference in reflection over a large spectral bandwidth. By employing this novel concept, we demonstrate reflectionless ($\sim$90\% transmission) half-wave plates, quarter-wave plates, and vector beam $q$-plates that can operate across multiple telecom bands with $\sim$99\% polarization conversion efficiency.
\end{abstract}

\maketitle

The concept of metamaterials -- artificial electromagnetic structures with unconventional properties composed of subwavelength constituent elements -- has opened many new opportunities for the efficient manipulation of light propagation and a design of novel types of metadevices with unusual properties~\cite{zheludev2012metamaterials}. In particular, {\em metasurfaces}, ultra-thin nanostructured layers, emerged recently as novel metadevices capable of reshaping transmitted and reflected light. Metasurfaces are composed of resonant subwavelength elements that are distributed spatially across a two-dimensional surface. Due to the resonant scattering of the incoming light, each element can alter the phase and amplitude of the light field. Metasurfaces thus enabled a variety of different functionalities realized at the subwavelength scale~\cite{Yu2014, minovich2015functional}, including controlled beam deflection, flat-lens focusing, and holography. As such, metasurfaces can replace the conventional free-space bulky components for wavefront and polarization control, imaging, and even on-chip integration.

While many designs and functionalities of metasurfaces have been suggested based on plasmonic planar structures~\cite{Yu2014, minovich2015functional}, most of these metasurfaces demonstrate low efficiencies in transmission due to losses in their metallic components.
Indeed, the maximum efficiency reported so far does not exceed 10$\%$~\cite{ni2013ultra}, and usually much lower efficiencies are reported (see Refs.~\cite{aieta2012aberration,walther2012spatial} to cite a few). In contrast, {\em all-dielectric resonant metasurfaces} completely avoid absorption losses, and can drastically enhance the overall efficiency~\cite{yang2014all, lin2014dielectric,decker2015high,arbabi2015subwavelength, arbabi2015dielectric, shalaev2015high},  especially in the transmission regime.
However, there exists a {\em serious issue} that prevents the immediate use of all-dielectric metasurfaces in practical planar metadevices: the fixed wavelength~\cite{Lalanne:1998:OL, arbabi2015dielectric} or narrow spectral range~\cite{decker2015high} of operation. Broadband and highly-efficient functionality was demonstrated for metasurfaces operating in reflection~\cite{yang2014dielectric}, but to date it has not been achieved for transmissive devices. Recently, transmissive metasurfaces operating at several different frequencies~\cite{aieta2015multiwavelength,2016arXiv160105847A} were demonstrated. Multiple frequency operation resolves the issue of narrow-band response for many specific applications, such as multi-colour imaging. However, for a number of technological implementations of metasurfaces , e.g. for polarization control of light, a continuous spectral range of operation is required, accompanied by clear design principles that can enable such broad bandwidth.

\begin{figure*}
\includegraphics[width=0.8\textwidth]{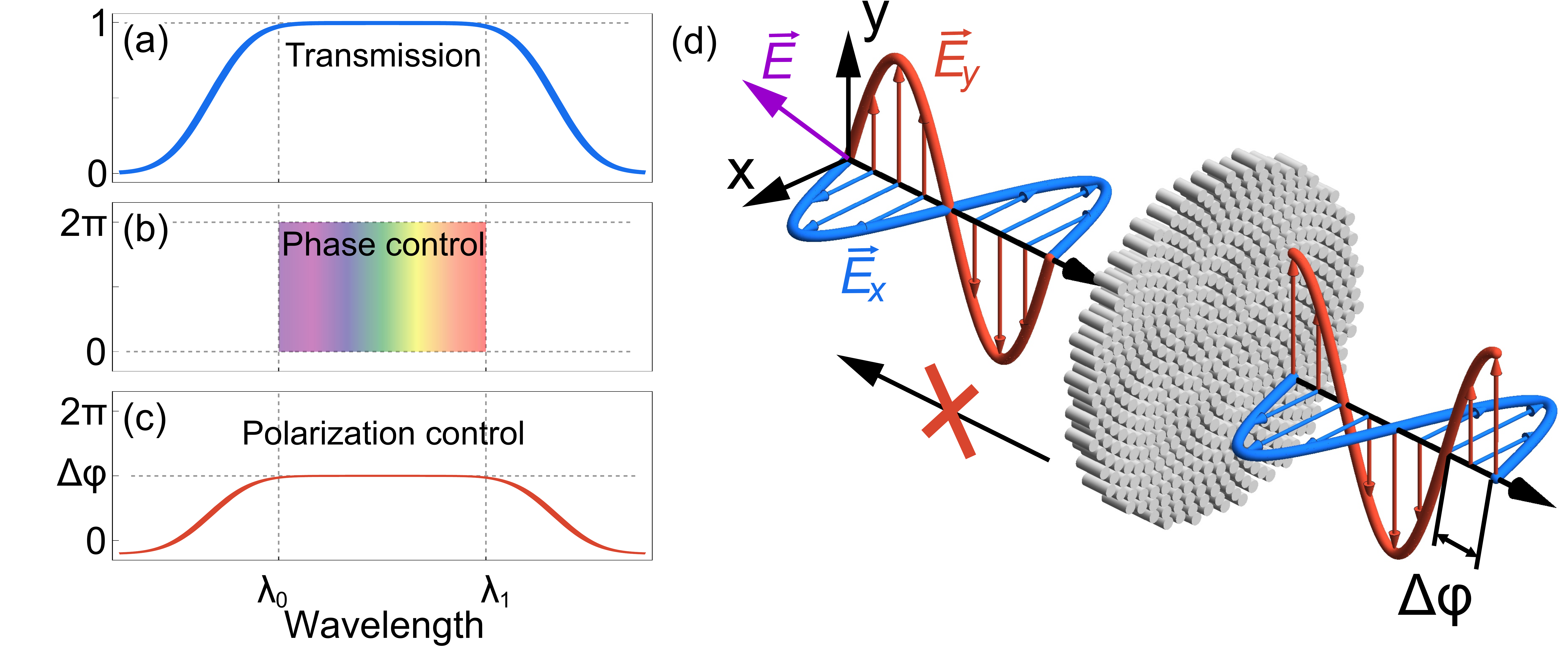}
\caption{\textbf{Concept of a broadband highly-efficient metadevice.} (a) Near-unity energy efficiency over a broad spectral region. (b) Phase coverage in the range [0 - 2$\pi$]. (c) Phase difference for two orthogonal polarizations in the range [0 - $\pi$]. (d) An artistic view of
 a metadevice for spatially-variant polarization control operating in a broad spectral range.}
\label{fig:Fig_Concept}
\end{figure*}

In this Letter, we suggest and demonstrate a novel approach for the design of broadband all-dielectric metasurfaces by overlapping the scattering contributions of several multipoles (both electric and magnetic) of the constituent nanoscale meta-atoms, thus producing high-transmission polarization and phase control over a wide spectral bandwidth. We employ this approach to demonstrate planar metadevices for efficient polarisation control, including half- or quarter-wave plates and vector beam $q$-plates. Our metadevices demonstrate near-100$\%$ energy efficiency in full-wave numerical simulations. 100$\%$ efficiency of a single-layer functional device is an advantage of employing resonant nanostructure as this allows for impedance-matching with air, which cannot be achieved with single-layer non-resonant functional structures. In experiment, the transmission efficiencies reach 90$\%$ and polarization conversion efficiency of $\sim$99\% across several telecommunication bands. While experimental energy efficiency is lower than theoretically predicted due to fabrication imperfections, it is among the highest demonstrated efficiencies of the metadevices. When compared to conventional birefringent media our devices demonstrate effective values of birefringence that are at least three times greater, and they allow for transverse patterning with pixel density at least three orders of magnitude better than any available liquid-crystal arrays.

To investigate the general operation of high-efficiency ultra-thin transmissive metadevices, we first review the basic conditions of operation for phase and polarization control.
\begin{enumerate}
\parskip0in
\itemsep0ex
\item{The device should provide high transmission, as illustrated in Fig.~\ref{fig:Fig_Concept}(a): $T\;\rightarrow\;100\%$.}
\item{The device should provide $2\pi$ phase control of transmission, relative to incident wave, thus covering the entire phase space [Fig.~\ref{fig:Fig_Concept}(b)]: $\varphi\,\in\,[0,2\pi]$ }
\item{The device should provide $\pi$ control of the phase difference between the transmission of two orthogonal polarizations [Fig.~\ref{fig:Fig_Concept}(c)]: $\varphi_x-\varphi_y\,\in\,[0,\pi]$ }
\end{enumerate}
The former two conditions  provide a high-transmission phase control, while the latter condition is further necessary to provide a polarization control.
In conventional optics, the first condition of broadband anti-reflection device can be fulfilled by employing multi-wave interference techniques from multiple stacked layers. However, satisfying all three conditions in a single layer metasurface over a broad bandwidth appears as an impossible task.

To tackle this problem, we must revisit the scattering of the constituent elements of the metadevice from the perspective of {\em multipolar interference}. We notice that light scattered from any arbitrary object can be decomposed into multipoles that have even or odd parity under spatial inversion (see Supporting Information), thereby allowing us to express the scattered field $\mathbf{E}$ as a sum of an even component $\mathbf{E}_{even}$ and an odd component $\mathbf{E}_{odd}$. As such, the transmission $T$ and reflection $R$ of a normally-incident, co-polarized plane wave propagating in the $\mathbf{\hat{z}}$ direction, can be written as follows,
\begin{align}
T &= |1 + E_{even}(\mathbf{\hat{z}}) + E_{odd}(\mathbf{\hat{z}})|^2\;,\\
R &= | E_{even}(\mathbf{-\hat{z}}) + E_{odd}(-\mathbf{\hat{z}})|^2=| E_{even}(\mathbf{\hat{z}}) - E_{odd}(\mathbf{\hat{z}})|^2 \;, \label{eq:thisTEXT}
\end{align}
where we have defined the magnitude and phase of the incident field such that $E_{\scriptscriptstyle \mathrm{PW}}(\mathbf{\hat{z}})=1$, and have assumed negligible absorption. However, to further control the phase $\phi$ of reflectionless transmission we need to satisfy:
\begin{align}
[1 + E_{even}(\mathbf{\hat{z}}) + E_{odd}(\mathbf{\hat{z}})] = e^{i \phi}.
\end{align}
Yet, because the magnitude is positive definite in Eq.~(\ref{eq:thisTEXT}), reflection can vanish only when $E_{even}(\mathbf{\hat{z}})$ and $E_{odd}(\mathbf{\hat{z}})$ are equal to each other in both magnitude and phase. As such, at any phase $\phi$,  the full transmission is achieved when the individual nanostructures of a metadevice radiate as:
\begin{align}
{E_{even}(\mathbf{\hat{z}}) = E_{odd}(\mathbf{\hat{z}})} = \frac{1}{2} \left(e^{i \phi}-1 \right)\;. \label{eq:conditionTEXT}
\end{align}
Notably, this equation can be satisfied for a single wavelength with a minimum of two multipoles, each having opposite parity; such as an electric and a magnetic dipole resonance. This single-wavelength functionality has previously been demonstrated for Huygens' metasurfaces~\cite{Pfeiffer:2013:PRL, Staude2013, decker2015high}. In particular, optical all-dielectric Huygens' metasurfaces rely on an overlap of the electric and magnetic resonances in high-index dielectric nanoparticles, resulting in zero backward scattering~\cite{Person2013, Fu:2013:NComm}. The idea of Huygens metasurface has proven successful in designing different functional metadevices, including vortex beam generators~\cite{Chong:2015:NL, shalaev2015high}, beam deflectors~\cite{Yu:2015:LPR, shalaev2015high} and holograms~\cite{arbabi2015dielectric, Chong2016}. However, since the overlap of the electric and magnetic dipole resonances (the so-called Huygens' condition) occurs only in a narrow spectral range, it is believed that no broadband devices can de achieved in this way.

To satisfy Eq.~(\ref{eq:conditionTEXT}) over a broad spectral range we instead employ a multipolar nanostructure, which makes $E_{even}$ and $E_{odd}$ each a linear combination of many multipoles. Importantly, Eq.~(\ref{eq:conditionTEXT}) then becomes a balance between multipoles, and does not require resonances of equal amplitude and spectral width. In other words, {\em multipolar structures} are inherently suited for realizing broadband phase control. Additional discussion regarding this conclusion and its derivation is provided in Supporting Information.

We can thereby achieve a broadband response as a result of multi-wave interference between the scattering waves produced by several pairs of multipoles of opposite parity. We refer to such multipolar forward scattering as the {\it generalized Huygens condition}, which resembles to an extent the forward scattering from two higher order multipoles called generalized Kerker condition~\cite{Liu:2014:OE, Alaee:2015:OL}. Such multipolar scattering also resembles some principles of operation from  multilayer dielectric coatings, where broadband functionality is a result of interference of multiple reflected waves.

\begin{figure}
\includegraphics[width=0.99\columnwidth]{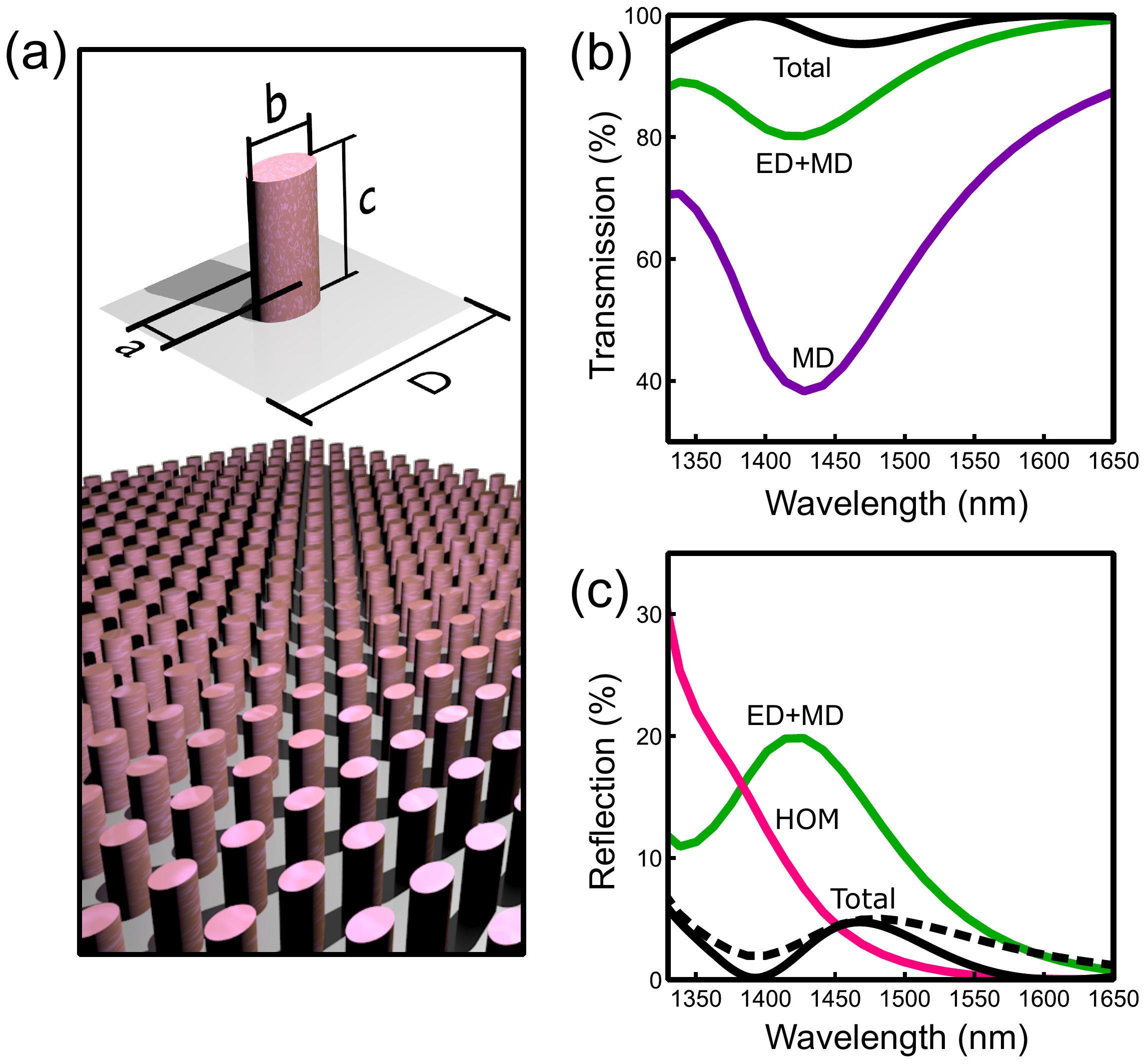}
\caption{(a) Schematic of our broadband metadevice. The spectra in (b) and (c) show the transmission and reflection of the metadevice and their dependence on the multipoles (ED = electric dipole, MD = magnetic dipole, and HOM = sum of higher-order multipoles). The black dashed curve in (c) shows the complete multipolar reflection when excluding reflection off the glass substrate.  Dimensions are $a = 285~\mathrm{nm},\; b = 450~\mathrm{nm}, \; c =860~\mathrm{nm}  \; \mathrm{and}\;D=750~\mathrm{nm}$, and the incident polarization is parallel to the short axis of the nanopillars.
}
\label{fig:2}
\end{figure}

We now focus on demonstrating the capacity of multipolar dielectric nanostructures for constructing broadband metasurfaces for highly-efficient polarization control.
Our metasurface consists of closely spaced silicon nanopillars, as shown in Fig.~\ref{fig:2}(a) to allow more multipolar modes in comparison with nanodisks.
Here we assume negligible optical losses and focus on minimizing reflection.
First, we relate the metasurface reflection to the magnitude of backward scattering from a single nanopillar with a superposition of individual contributions of higher-order multipoles [details in Supporting Information].
As seen in Fig.~\ref{fig:2}(b), the silicon nanopillar array approaches broadband near-unity transmission as we account for higher-order multipoles.
 In Fig.~\ref{fig:2}(c), we show that the origin of this broadband transmission is attributed to interference in the reflection  between dipole- and higher-order multipoles (HOM).
Both dipole-order multipoles and the higher-order multipoles can be seen to provide a significant contribution to reflection in isolation, while their sum suppresses reflection.
Because of the negligible material losses of silicon, this broad suppression of reflection is sufficient to provide near-unity and broadband transmission through the metasurface.

\begin{figure}
\includegraphics[width=0.8\columnwidth]{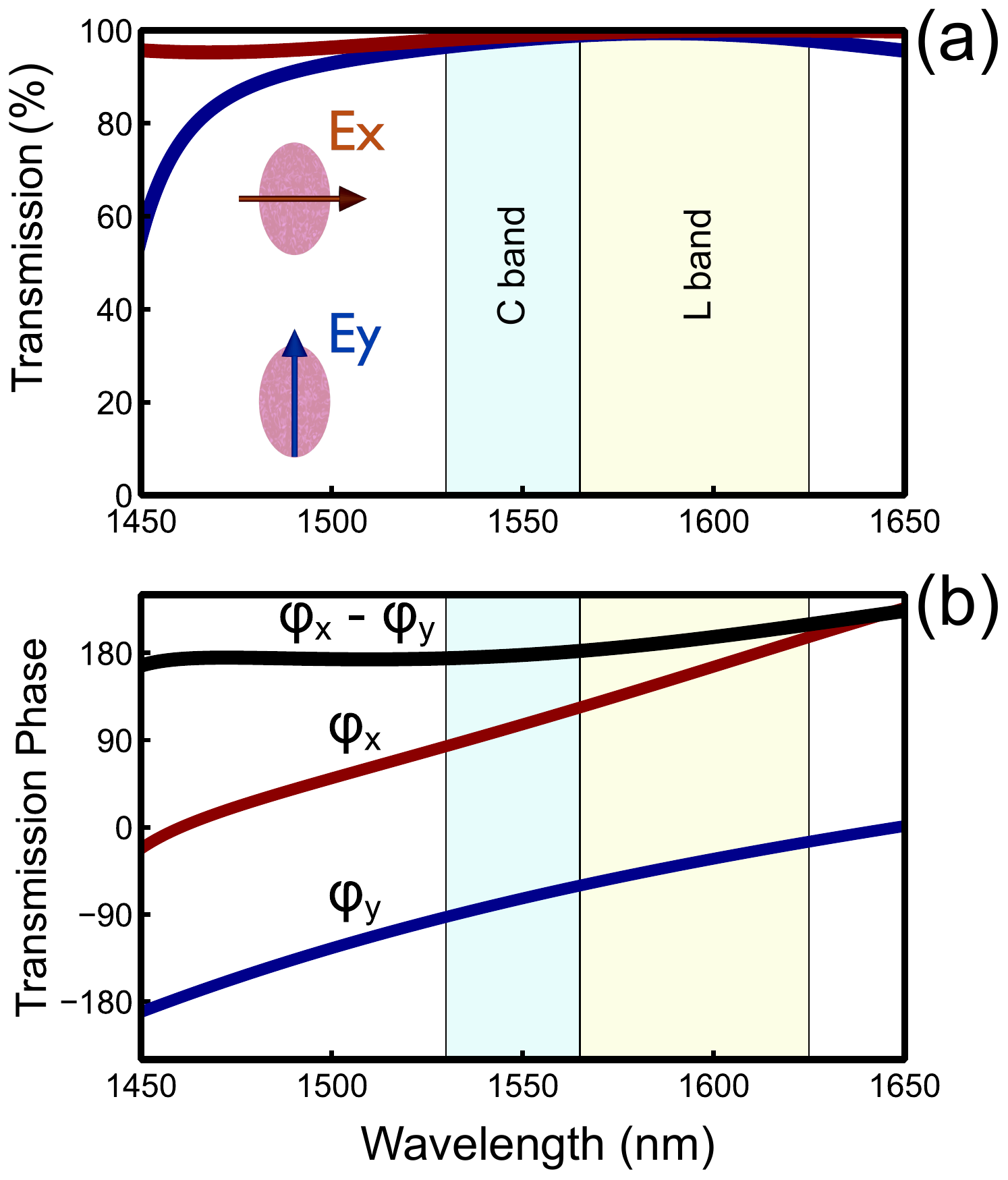}
\caption{(a) Transmission spectrum of the metadevice shown in Fig.~\ref{fig:2}, for polarizations parallel to the two axes of the nanopillar. (b) Phase of the transmitted waves with respect to the incident field for both orthogonal polarizations. The phase difference $\varphi_x-\varphi_y$ shows a constant value of $\pi$ over a broad spectral range, as required for a broadband half-wave plate.}
\label{fig:3}
\end{figure}

As the next step, we need to achieve the control over the phase of the transmitted wave. In this regard, it is important to recognize that the forward scattering components from each active multipole must collectively suppress the incident field and construct a new wavefront that can have a different phase to that of the background field. Therefore, the scattering contributions from the various multipoles have to be non-negligible relative to the incident field. This is indeed the situation in Figs.~\ref{fig:2}(b,c), where both the dipole- and the higher-order multipoles are shown to provide a significant contribution to transmission and reflection in isolation.

To finally demonstrate the combined phase and polarization control of the transmitted wavefront, we can maintain a near constant relative phase difference between the transmitted wave polarized along each axis of the silicon nanopillars of Fig.~\ref{fig:2}(a). In particular, we consider the case of a half-wave plate in Fig.~\ref{fig:3}, and maintain a $\pi$ phase difference between the transmission polarized along each axis. Importantly, our results show that the $\pi$ phase difference is maintained together with near unity transmission over both C and L telecommunications bands.

\begin{figure*}
\includegraphics[width=0.75\textwidth]{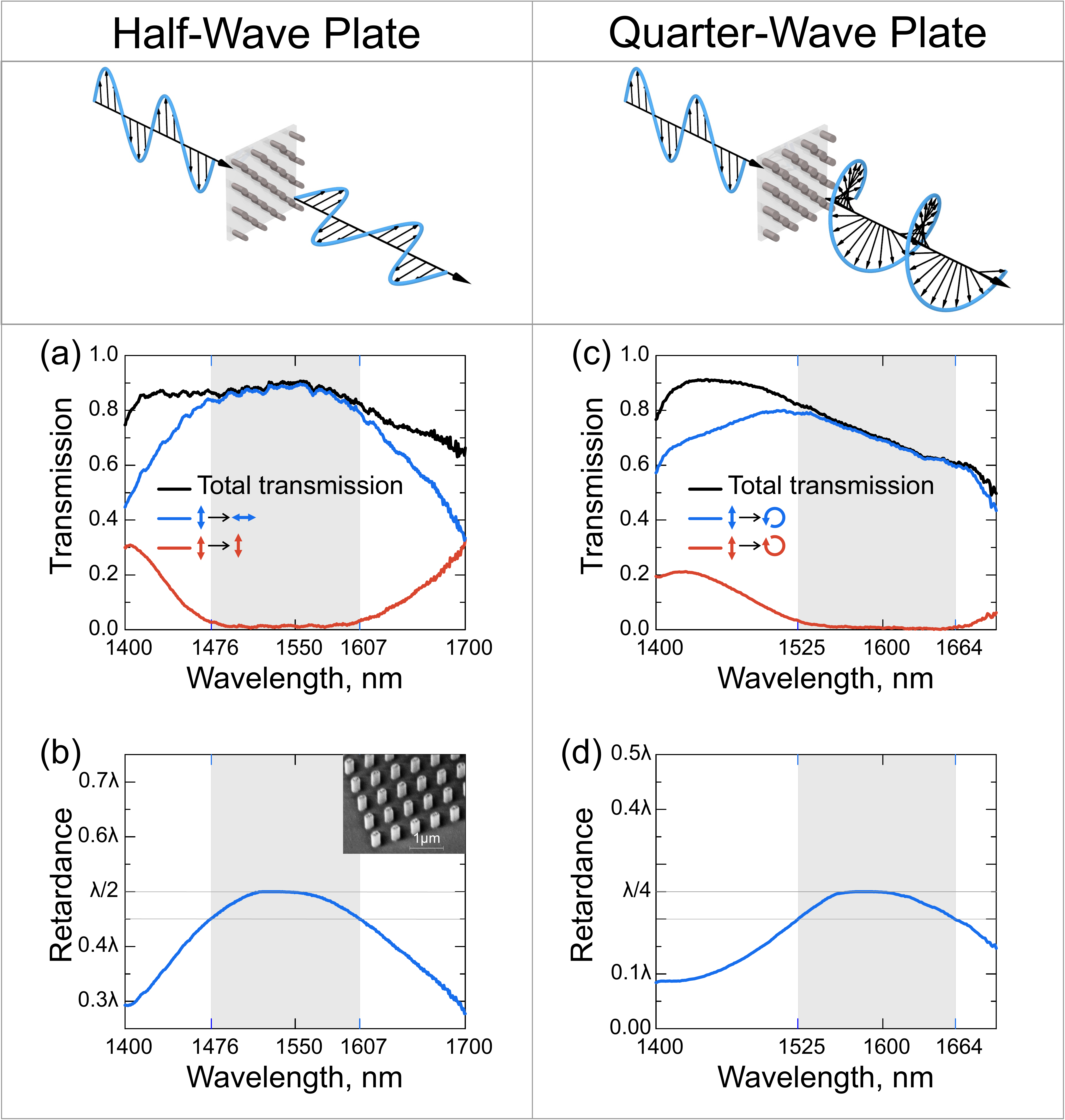}
\caption{\textbf{Demonstration of half- and quarter-wave plates.} (a) Experimental transmission spectra of the half-wave plate. Black line -- \emph{total transmission}. Red line -- transmission through a metasurface positioned between a pair of linear \emph{parallel polarizer and analyzer} with polarization axis oriented at 45$^\circ$ to metasurface's anisotropy axis. Blue line -- transmission through a metasurface positioned between a pair of linear \emph{crossed polarizer and analyzer} with polarization axis oriented at $\pm$45$^\circ$ to metasurface's anisotropy axis. (b) Experimentally retrieved retardance of the half-wave plate. Highlighted region shows the spectral range, with a deviation from $\lambda/2$ by no more that 10$\%$, or $\lambda/20$. Inset shows a scanning electron micrograph of the fabricated half-wave plate. (c) Experimental transmission spectra of the quarter-wave plate. Black line -- \emph{total transmission}. Red line -- transmission through a metasurface positioned between a pair of linear polarizer with polarization axis oriented at 45$^\circ$ to metasurface's anisotropy axis and a right-circular analyzer. Blue line -- transmission through  a pair of linear polarizer with polarization axis oriented at 45$^\circ$ to metasurface's anisotropy axis and a left-circular analyzer. (d) Experimentally retrieved retardance of the quarter-wave plate. Highlighted region show the spectral range, with retardance deviation from $\lambda/4$ by no more than $\lambda/20$.}
\label{fig:Fig_Experiment_1}
\end{figure*}

High operation bandwidth is a result of multipolar interference  that can be achieved in various systems not limited to dielectric nanopillars. On the other hand, the use of similar nanopillar geometries may not meet  the conditions for the multipolar interference and thus lead to essentially single-wavelength operation. In Supporting Information we give an example of two metadevices  with similar geometries of nanopillars, but featuring 8 times difference in the spectral bandwidth.

\begin{figure*}
\includegraphics[width=0.6\textwidth]{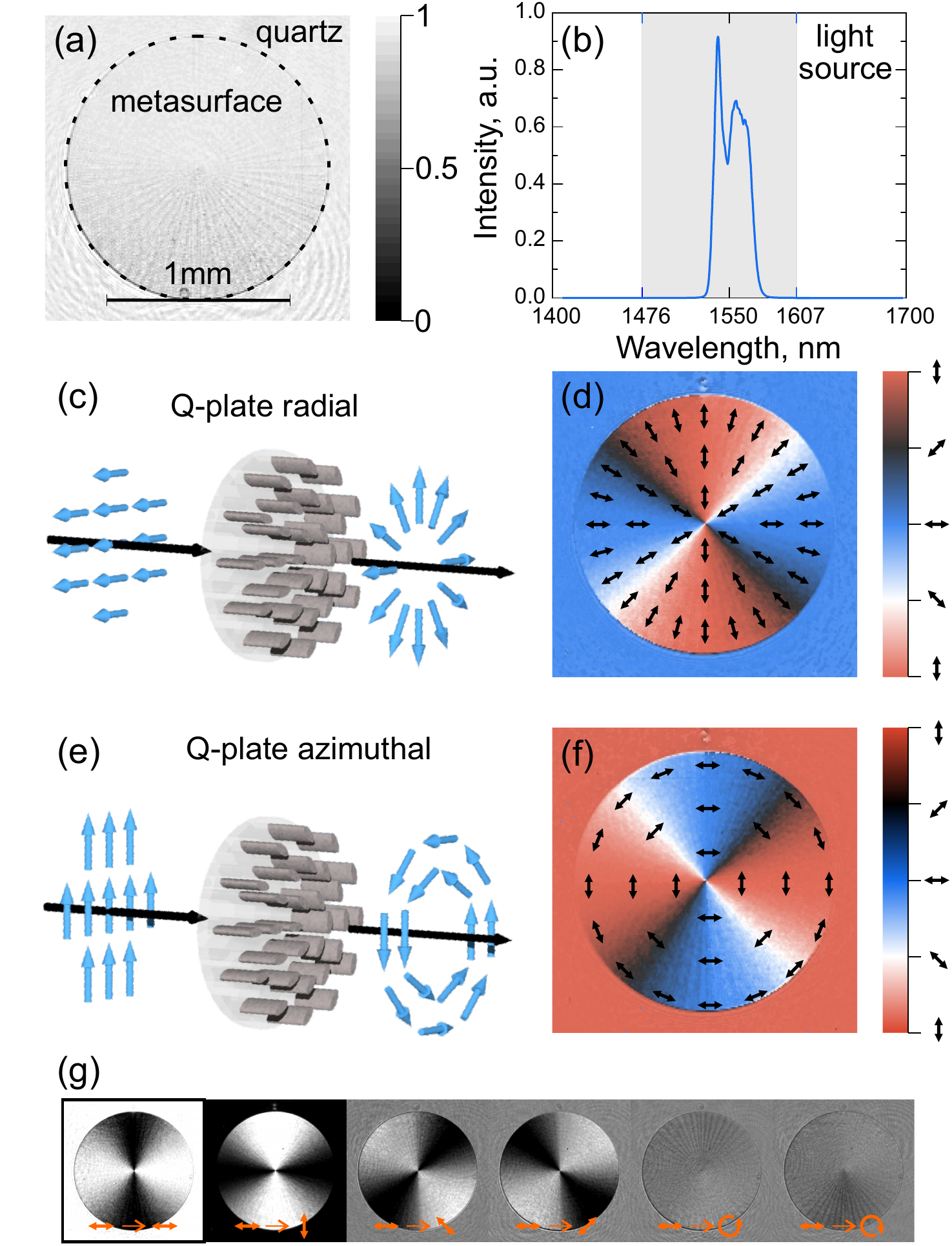}
\caption{\textbf{Demonstration of highly-efficient $q$-plates.} (a) Transmission image of the $q$-plate ~1.5 mm in diameter placed on top of a quartz wafer, taken with an amplified spontaneous emission (ASE) broadband light source at around 1550\,nm wavelengths. Note that the device appear to be nearly transparent. (b) Spectrum of the ASE light source used for the experiment. (c) $q$-plate functionality concept: conversion of vertical into azimuthal polarization. (d) Experimentally measured spatially-resolved polarization of the azimuthally polarized  output beam. (e) The $q$-plate functionality concept:  conversion of horizontal into radial polarization.  (f) Experimentally measured spatially-resolved polarization of the radially polarized output beam. (g) Examples of light transmission through the $q$-plate positioned between different polarizers and analyzers (polarizer axes orientations marked in insets).}
\label{fig:Fig_Experiment_2}
\end{figure*}

We then proceed to experimental realization and optical testing of the proposed broadband functionality. We fabricate our samples from polycrystalline silicon on quartz substrate using electron-beam lithography [see details in Supporting Information]. Overall, we demonstrate {\em three types of optical metadevices} -- a half-wave plate, a quarter-wave plate, and a $q$-plate all working in the telecom spectral region. The half-wave plate is an element that adds $\pi$ phase difference between two principal linear polarizations of light. It is used for polarization rotation by up-to 90$^o$. Our half-wave plate consists of nanoparticles with sizes $a = 280~\mathrm{nm},\; b = 430~\mathrm{nm}, \; c =850~\mathrm{nm}$, arranged into a square lattice with period $D=750$\,nm. The $q$-plates are used for conversion of linearly-polarized beam into radially, or azimuthally polarized. Our $q$-plate consists of the same nanoparticles arranged into hexagonal lattice with the same density of particles per unit area. The spatial orientation of the nanoparticles, however varies spatially across the $q$-plate, as shown in Fig.~\ref{fig:Fig_Experiment_2}. The quarter-wave plate is an element that adds $\pi/2$ phase difference between two principal linear polarizations of light. It is used for polarization conversions between linear, circular and intermediate elliptical polarizations. Our half-wave plate consists of nanoparticles with sizes $a = 310~\mathrm{nm},\; b = 370~\mathrm{nm}$, same height $c=850~\mathrm{nm}$ and arranged into the same square lattice ($D=750~\mathrm{nm}$).

We then study experimentally the optical properties of the fabricated wave-plates. First, we measure their transmission spectra. For this, we employ a home-built white-light spectrometer that uses a tungsten light bulb as a broadband optical source and an Optical Spectrum Analyzer detection. The transmission spectra of the samples
are shown in Figs.~\ref{fig:Fig_Experiment_1}(a,c--black lines).

To check the basic functionalities of the metadevices, we measure how well the half-wave platecan rotate linear polarization by $90^\circ$ and how well the quarter-wave plate can convert linear polarization into circular one. We start with the fabricated half-wave plate and place it between linear polarizer and analyzer with the wave plate axis being at $45^\circ$ to the axis of the polarizer (the wave plate axis is defined here as the long axis of constituent elliptical nanoparticles). We observe that in this case the half-wave plate rotates linear polarization of virtually $100\%$ of transmitted light by $90^\circ$ over a broad spectral region (shown as shaded region in the plots). In particular, Fig.~\ref{fig:Fig_Experiment_1}(a) shows that transmission through a wave plate positioned between two crossed polarizers is reaching its maximum value, and transmission through a wave plate positioned between two parallel polarizers almost vanishes.

Similarly, we test the performance of the quarter-wave plate. We place a linear polarizer in front of it and orient the wave plate axis at $45^\circ$ with respect to the polarizer's axis. After the wave plate we place a circular polarizer (transmitting either left- or right-circular polarization) made of a superachromatic quarter-wave plate (from B. Halle) and a linear analyzer. We demonstrate virtually $100\%$ conversion of linear polarization into circular over a broad spectral region, as shown in Fig.~\ref{fig:Fig_Experiment_1}(c).

We further perform the full polarimetry measurements of the transmission spectra of the metasurfaces. For this we keep a vertically aligned linear polarizer and orient the samples anisotropy axis at $45^\circ$. We use a set of six different analyzers (six different orientations of the super-achromatic quarter-wave plate and a linear analyzer): linear horizontal and vertical, two linear diagonal, left-circular and right-circular analyzers; and correspondingly record six spectra. We then use the Stokes vector formalism to retrieve the full information about the polarization states of the transmission spectra in terms of their polarization degree, ellipticity and polarization rotation angle [see more details in Supporting Information]. The extracted retardance of our wave plates is shown in Figs.~\ref{fig:Fig_Experiment_1}(b,d). Clearly, the measured retardance is close to 0.5 for the case of the half-wave plate and 0.25 for the case of the quarter-wave plate. Importantly, these values are constant (within the practically accepted tolerances) over a broad spectral range.

The biggest significance of the metasurfaces approach however comes from the ability to spatially vary the degree of polarization rotation. We therefore demonstrate experimentally the broadband operation of a $q$-plate, which is a device that converts beam with linear vertical polarization into a beam with azimuthal polarization and a beam with horizontal polarization into a beam with radial polarization [see Fig.~\ref{fig:Fig_Experiment_2}]. We build an image of the $q$-plate on an infrared Xenics camera and use an amplified spontaneous emission broadband light source such that on the camera we receive an image [see Fig.~\ref{fig:Fig_Experiment_2}(a)] \emph{integrated} over the entire spectral range of our source, shown in Fig.~\ref{fig:Fig_Experiment_2}(b). The $q$-plate shows $\sim90\%$ overall transmission and it looks nearly transparent on the camera. Similarly to the studies of the wave plates, we perform the full polarimetry of the optical images [see more details in Supporting Information], i.e. we capture six images of the $q$-plate when the light passes through six different analyzers [see Fig.~\ref{fig:Fig_Experiment_2}(g)]. We then employ the Stokes vector formalism and, in particular, retrieve polarization angles of light that passes through different spatial points of the $q$-plate [see Figs.~\ref{fig:Fig_Experiment_2}(d,f), different colors correspond to different directions of linear polarization]. Figures~\ref{fig:Fig_Experiment_2}(d,f) show that the metadevice allows virtually perfect conversion of vertical polarization into radial polarization and horizontal polarization into azimuthal polarization.

In conclusion, we have designed and demonstrated transparent all-dielectric metadevices for highly efficient polarization manipulation based on the generalized Huygens' principle, including half-wave plates, quarter-wave plates, and vector beam $q$-plates that can operate across several telecom bands. While the demonstrated metadevices allow to change the polarization of light in a manner similar to conventional birefringent crystals and liquid crystals, the effective value of birefringence is much higher than that in conventional materials. Indeed, for both half-wave plates and $q$-plates, the $\pi$ phase difference is accumulated over $850$\,nm distance at $1550$\,nm wavelength, being equivalent to $\Delta n=(n_o-n_e)\approx0.9$. The highest values of $\Delta n$ for conventional materials are of the order of 0.3~\cite{devore1951refractive, li2005infrared}. The metadevices feature subdiffractional spatial resolution of parameters variation, which is  three orders of magnitude higher than that in the state-of-the-art liquid crystal arrays. The spectral bandwidth of the demonstrated metadevices is comparable or larger than the bandwidth of conventional zero-order wave retarders and diffractive optical elements.

\section*{Acknowledgements}
The authors acknowledge a support of the Australian Research Council and thank F. Capasso, T. Krauss, and S. Turitsyn for stimulating discussions.  A portion of this research was conducted at the Center for Nanophase Materials Sciences sponsored at Oak Ridge National Laboratory by the Scientific User Facilities Division, Office of Basic Energy Sciences, the US Department of Energy.

\end{document}